\documentclass[12pt]{iopart}
\usepackage{iopams}  
\usepackage[dvips]{graphicx}
\usepackage{color}
\usepackage{subfigure}
\definecolor{mycolor}{rgb}{1,0,0}

\begin{document}
\title[Stuttering Min oscillations within {\it E. coli}: A stochastic polymerization model]{Stuttering Min oscillations within {\it E. coli} bacteria: a stochastic polymerization model} 

\author{Supratim Sengupta}
\address{Centre for Computational Biology and Bioinformatics \\ 
School of Computational \& Integrative Sciences \\ 
Jawaharlal Nehru University, New Delhi-110 067, India.\\
and \\
Department of Physical Sciences \\
Indian Institute of Science Education \& Research, Kolkata \\
Mohanpur Campus, Nadia, West Bengal - 741252, India
}
\ead{supratim.sen@iiserkol.ac.in}

\author{Julien Derr}
\address{Laboratoire Mati\`{e}re et Syst\`{e}mes Complexes, Universit\'{e} Paris Diderot, 10 rue Alice Domont et L\'{e}onie Duquet, F-75205 Paris cedex 13, France} 
\ead{julien.derr@univ-paris-diderot.fr}

\author{Anirban Sain}
\address{Department of Physics, Indian Institute of 
Technology-Bombay, Powai, Mumbai, 400076, India}
\ead{asain@phy.iitb.ac.in}

\author{Andrew D Rutenberg}
\address{Department of Physics and Atmospheric 
Science, Dalhousie University, Halifax, Nova Scotia, Canada, B3H 4R2}
\ead{adr@dal.ca}

\vfil
\vspace*{1cm}

\today 
\submitto{\PB} 

\pagebreak
%%%%%%%%%%%%%%%%%%%%%%%%
\begin{abstract}
We have developed a 3D off-lattice stochastic polymerization model to study subcellular oscillation of Min proteins in the bacteria {\em Escherichia coli}, and used it to  investigate the experimental phenomenon of Min oscillation stuttering.  Stuttering was affected by the rate of immediate rebinding of MinE released from depolymerizing filament tips (processivity),  protection of depolymerizing filament tips from MinD binding, and  fragmentation of MinD filaments due to MinE.   Each of processivity, protection, and fragmentation reduces stuttering, speeds oscillations, and reduces MinD filament lengths.  Neither processivity or tip-protection were, on their own, sufficient to produce fast stutter-free oscillations. While filament fragmentation could, on its own, lead to fast oscillations with infrequent stuttering; high levels of fragmentation degraded oscillations.  The infrequent stuttering observed in standard Min oscillations are consistent with short filaments of MinD, while we expect that mutants that exhibit higher stuttering frequencies will exhibit longer MinD filaments. Increased stuttering rate may be a useful diagnostic to find observable MinD polymerization in experimental conditions. 
\end{abstract}

\maketitle

%%%%%%%%%%%%%
\section{Introduction}

Subcellular oscillations of the proteins MinD and MinE within the rod-shaped bacterium {\it E. coli} help restrict division to midcell \cite{Raskin1997, Raskin1999, Lutkenhaus1999, Rothfield2005, Kruse2005, Lutkenhaus2007}. Pole-to-pole Min oscillations arise from an  interplay between diffusion and membrane binding/unbinding of MinD and MinE proteins in the confined bacterial geometry. MinD-ATP binds to the membrane in a cooperative manner, and MinE  binds to the membrane-bound MinD-ATP.  Subsequent MinE-stimulated hydrolysis of membrane-bound MinD-ATP leads to the release of the MinD-ADP into the cytoplasm, together with MinE. Released MinE can immediately rebind to nearby membrane bound MinD-ATP, but MinD-ADP must undergo nucleotide exchange before it can rebind. In rod-shaped {\em E. coli}, membrane associated MinD is observed to form polar caps at alternating poles, and MinE associates with the medial edge of these caps in ring-like structure (the ``E-ring''). 

Dynamic filamentous structures of MinD have also been reported in vivo \cite{Shih2003,Szeto2005}, and are consistent with the observation of MinD polymerization in vitro \cite{Hu2002,Suefuji2002} and the delays of MinE-stimulated MinD-ATPase activity seen {\it in vitro} \cite{Derr2009}.  The Min oscillation {\it in vivo} may thus involve the periodic polar nucleation, polymerization, and depolymerization of MinD filaments.  Nevertheless, long polymeric structures are not seen in electron cryotomograms \cite{Swulius2011}, and while static filamentous MinD structures were reported in a reconstituted planar Min system \cite{Loose2008} (see Fig.~S11) dynamic filaments were not reported. It remains unclear how ubiquitous membrane associated MinD polymers are in normal {\it  E.coli} Min oscillations \cite{Cabeen2010,Loose2011}, and what length any polymeric filaments have. While short MinD polymers would not be easily observable, could they still significantly affect the observable phenotype of the Min oscillation?

The stuttering of the disassembly of the polar MinD caps that has been observed in wild-type (WT) Min oscillations \cite{Hale2001EMBO} and that is common in certain mutant systems \cite{Shih2002EMBO, Hsieh2010} has not been recovered in existing models of Min oscillation \cite{Fange-Elf2006}.  Organization of a small number of Min molecules (roughly 2000 MinD monomers and 700 MinE dimers \cite{Shih2002EMBO}) into an even smaller number of filaments should enhance stochastic effects. Indeed, Min stuttering has a natural explanation in tip-directed depolymerization models, where the bistability of individual filaments selected by MinE tip-decoration allows switching of individual filaments between disassembly and growth \cite{Derr2009}. If many of the filament tips are denuded of MinE at the same time, which switches them from depolymerization to polymerization, then the collective oscillation should stutter.  

Surprisingly, existing stochastic polymerization models \cite{Drew2005, Pavin2006, Krstic2007, Tostevin2006, Cytrynbaum2009} have not reproduced the stuttering phenomenon.  Instead, we believe that stuttering has been strongly suppressed in all existing models. The stochastically-switched  1d model of Borowski and Cytrynbaum \cite{Cytrynbaum2009} does not allow switching while the filament tip is decorated with any MinE and does not have any spatial distribution of free monomers, and so cannot capture temporary reversal i.e. stuttering. Similarly, the 1d  model of Drew {\em et al.} \cite{Drew2005} precludes temporary reversals by keeping all of the bound MinE at the MinD-tip.  Alternatively, the 1d model of Tostevin and Howard \cite{Tostevin2006} allows MinE-induced filament fragmentation or cutting by allowing slow MinD unbinding away from filament tips. The resulting proliferation of rapidly-depolymerizing tips is similar to what is seen in non-polymerizing reaction-diffusion models. This proliferation of tips avoids collective poisoning. A similar mechanism of filament fragmentation was used by Pavin {\em et al.} \cite{Pavin2006} and Krsti\'{c} {\em et al.} \cite{Krstic2007} in 3D. 

In this paper we explore three tuneable mechanisms to inhibit poisoning of individual filaments in a polymerizing model for Min oscillations. The first is to allow for processivity of MinE (with the parameter $P_{passE}$, described below), such that tip-bound MinE is not always released from the filament upon depolymerization but has a chance to immediately rebind \cite{Derr2009, Klein2005}.  This is similar to the ``Tarzan of the Jungle'' mechanism proposed by Park {\em et al} \cite{Park2011}, but applied to polymeric MinD.  However, we find that this mechanism is not enough on its own to suppress stuttering. So in addition to processivity, our second  mechanism is to allow for tip-bound MinE to protect the filament from further MinD-ATP binding (with the parameter $P_{protect}$, described below). This protection mechanism was also included in the polymeric model of Tostevin and Howard \cite{Tostevin2006}. The third mechanism is to allow MinE that are bound to MinD filaments to cut filaments away from the tip (with rate $k_f$, described below) \cite{Pavin2006, Krstic2007, Tostevin2006}.   

We investigate stuttering, or transient reversals of the disassembly of polar MinD caps, within the context of a stochastic polymerization model of the Min oscillation. This is a 3D, molecular-dynamics style, ``all-molecules'' model, where the stochastic effects due to shot-noise and stochastic binding and unbinding are all implicitly included. 

%%%%%%%%%%%%%%%%%%%%%%%%%
\section {Our Stochastic Polymerization Model}

We model the 3D {\it E. coli} bacterium by a cylindrical membrane of length $2L$ and radius $R$, capped with two polar hemispheres of radius $R$ --- all enclosing the fluid bacterial cytoplasm where diffusion occurs. We work with experimental number densities  scaled to an {\it E. coli} cell of length $4\mu m$ and diameter $1\mu m$ and so containing approximately $3500$ MinD monomers and $1200$ MinE dimers \cite{Shih2002EMBO}. Each cytoplasmic MinD or MinE diffuses by taking a randomly-oriented (isotropic) step of fixed length $\delta$ in every timestep $\Delta t$, leading to a diffusion constant $D=\delta^2/(6\Delta t)$. We use $D_D=16\mu m^2/s$ and $D_E = 10\mu m^2/s$ for MinD and MinE \cite{Meacci2006}, respectively, and choose $\Delta t=10^{-2}s$.  Both MinD monomers and MinE dimers are treated as non-interacting particles while diffusing.  

MinD-ATP monomers can bind as an isolated monomer (with rate $\sigma_D$, see Fig.~1(a)) or can bind to an isolated bound MinD monomer (with rate $\sigma_{nuc}$, see Fig.~1(b)). These rates naturally translate into probabilities (see next subsection) for diffusing particles that encounter the membrane. Cooperative binding to a bound MinD monomer initializes a polymer of length $2 a_0$, where $a_0 = 5nm$ is the bound subunit spacing along MinD protofilaments  \cite{Suefuji2002}. The orientation of the growing polymer is determined at nucleation. In this paper, we have taken all polymers to be straight along the geodesic line from one cell pole to the other: i.e. on great circles on end-caps and axial along the cylindrical portion of the membrane.  MinD-ATP monomers can extend an existing polymer by binding to its tip monomer with a rate $\sigma_{dD}$ (see Fig.~1(c)). Every monomer added extends the polymer by $a_0$.  MinD-ATP binds to the closest membrane-bound MinD, either monomer with rate $\sigma_{nuc}$ or tip with rate $\sigma_{dD}$, that lies within the radial distance $r_{nuc}$ or $r_D$, respectively, of the point where MinD strikes the membrane.  We require this finite interaction range in order to speed our computational algorithm: it allows us to take larger random steps during diffusion, and it allows us to turn off membrane diffusivity of MinD. Since most MinD is membrane associated, this represents an enormous computational efficiency.

To include the effect of PL heterogeneity \cite{EugeniaCLPatch,EugeniaReview} and account for the observation that MinD associates with anionic CL-rich PL more than non-polar PL \cite{EugeniaDeltaPE}, we have allowed the rate ($\sigma_{Dcl}$) of association of MinD to the CL-rich end-caps to be greater than its rate ($\sigma_{D}$) of association elsewhere on the membrane.  We furthermore only allow MinD filament nucleation (via $\sigma_{nuc}$) only at the CL-rich polar caps.   We also briefly consider a fully homogeneous model with $\sigma_{Dcl}=\sigma_D$ and homogeneous nucleation.

MinE can bind (with rate $\sigma_E$, see Fig.~1(d)) to the closest membrane bound MinD-ATP that lies within distance $r_E$ of the point where MinE strikes the membrane. We report results for  $r_E = r_D = 5 a_0$, and $r_{nuc}=a_0$. However, we have explored a wide range of $r_D$ and $r_E$ values and observed Min oscillations for $r_D$ and $r_E$ values as low as $a_0$ and $2 a_0$ respectively.   MinE can bind to any membrane associated MinD, whether they are in filaments or not. When MinE binds, it forms a MinDE complex. Unbinding of MinDE primarily proceeds through MinE stimulated MinD ATPase activity of the MinDE complex. If the MinE is bound to an isolated membrane-bound MinD monomer the release rate is $k_{SM}$, the release rate at the tips of a filament is $k_S$ (see Fig.~1(e)).  We investigate the effects of MinE-stimulated MinD release away from filament tips (through the rate $k_f$, where $k_f \leq k_S$), which cuts or fragments the filament (see Fig.~1(h)). We also allow a small intrinsic (non-hydrolysed) release rate of  bound MinD monomers with rate $k_I$.   MinE can also spontaneously release from a MinDE complex without hydrolysis of the associated MinD with rate $k_E$. These small spontaneous release rates of MinD and MinE without hydrolysis represent the reversibility of binding interactions.   Spontaneously released MinD, like MinE,  need no recovery time before rebinding. In contrast, after stimulated release MinD-ADP spends time $\tau_c$ in the cytoplasm before it is converted to MinD-ATP by nucleotide exchange and become capable of binding to the membrane again \cite{Huang2003}.  

When MinE is released from the filament tip it is passed on to the next MinD site on the filament with a ``passing'', or processivity, probability $P_{passE}$, provided the site is not already occupied by a MinE --- see Fig.~1(f).  Processivity arises naturally from the efficient exploration of the tip-environment by a continuously diffusing tip-released MinE \cite{Derr2009}.  Processivity may also be enhanced by MinE-membrane interactions \cite{Park2011}. In addition to processivity, we inhibit binding of MinD to the tips of filaments that are already decorated by MinE with probability $P_{protect}$ (see Fig.~1(g)).   Both $P_{passE}$ and $P_{protect}$ range from $0$ to $1$, and serve as tuneable parameters that inhibit poisoning of individual filaments. 

%%%%%%%%%%%%%%%%%%%%%%%%%%%%%
\subsection {Implementation details}

As mentioned above, we implement diffusive motion through a randomly oriented step of fixed length $\delta_0$, where $D= \delta_0^2/(6 \Delta t_0)$. We keep the timestep $\Delta t_0$ fixed to allow synchronous motion, and so adjust the step size $\delta_0$ to give the desired cytoplasmic diffusivities of MinD and MinE.   If the step would cross the cytoplasmic membrane then binding is checked (see below). If the particle does not bind, it is reflected specularly from the membrane. This ensures a uniform volume density in the non-interacting limit. For computational efficiency we have chosen a relatively large maximal timestep $\Delta t_0=10^{-2} s$ that is still much less than the Min oscillation period. This timestep is used for unbinding, for nucleotide exchange, and for diffusing proteins far from the membrane. For bulk MinD-ATP or MinE close to the membrane, we use a random spatial step length $\delta$ equal to half-the separation of the protein from the closest membrane --- but no less then $a_0 = 5nm$.  We adjust the timestep $\Delta t = \delta^2/(6D)$ accordingly, and take these smaller steps until $\Delta t_0$ is reached.  As a result the simulation is efficient and synchronous, but retains a relatively fine spatial scale close to the membrane. 

Reaction-diffusion membrane association rates are mapped to ``sticking'' probabilities of MinD-ATP and MinE upon collision with the cytoplasmic membrane \cite{Derr2009}. The mapping agrees dimensionally with the one given by Pavin {\em et al.} \cite{Pavin2006,Krstic2007} but our dimensionless prefactors differ. For a particle a distance $z< \delta$ from a membrane, steps within a polar angle $\theta_z = \cos^{-1}(z/\delta)$ will hit the membrane. The corresponding solid angle gives a fraction $f(z)=(1-z/\delta)/2$ of particles hitting the surface with one randomly oriented step of length $\delta$. For a bulk density $\rho$, integrating over $z$ gives a sticking rate per unit area of $P \rho \delta/4$ --- where $P$ is the desired sticking probability. Equating this to the expected change in one timestep using reaction-diffusion rates, $\sigma \rho \Delta t$, gives $P =4 \sigma \Delta t/\delta = 2 \sigma \delta/(3D)$.  This is used for MinD-ATP binding to the membrane. For binding to a small patch of area $\pi r^2$, e.g. at the filament tip, we can use the previous rate per unit area to obtain the binding rate $ \pi r^2 P_{tip} \rho \delta/4$ and equate that to the reaction-diffusion rate $\sigma_{tip} \rho \Delta t$ to obtain $P_{tip} = 2 \sigma_{tip} \delta/ (3 \pi r^2 D)$. We use this for MinD-ATP binding to the filament tip, filament nucleation, and for MinE binding to membrane associated MinD.  We return particles released from the membrane to the position that they originally bound from in order to recover uniform bulk densities in the absence of other interactions.

We have investigated a variety of initial conditions, such as randomly distributed MinD and MinE or an inhomogeneous condition with all MinD randomly placed near one pole and all MinE randomly placed near the other. Our results are unaffected by the initial conditions, though we typically used inhomogeneous initial conditions to minimize the duration of initial transients before steady-state data could be taken.

%%%%%%
\subsection {Data analysis}
To simply characterize oscillations in a manner amenable to experimental measurement, we have recorded the amount of membrane-associated MinD, $n_D(t)$, in each polar hemispherical cap at $1s$ intervals. We have also recorded the number of filaments, and their length. The plots of MinD in one pole vs time, as illustrated in Figs.~2 and 3, showed an initial transient and then periodic oscillations. We ignored the earliest 10\% of the data (similar results were obtained with 20\%) to avoid initial transients. We then characterized the time-average $\langle n_D \rangle$, the variance $\sigma_D^2 = \langle n_D^2 \rangle - \langle n_D \rangle^2$, and the corresponding standard deviation $\sigma_D \equiv \sqrt{\sigma_D^2}$. 

From the $n_D(t)$ plot shown in Fig.~2, we identified local minima and maxima that were below $\langle n_D \rangle - \sigma_D$ or above $\langle n_D \rangle +\sigma_D$, respectively in order to accurately identify the peak and trough of the oscillations. These local minima or maxima sometimes clustered, but the clusters alternated between minima and maxima. Within each cluster we took the smallest minimum or largest maximum as the corresponding extremum of one oscillation. The region  $n_D \in \left[\langle n_D \rangle - \sigma_D, \langle n_D \rangle + \sigma_D \right]$, from one maximum to the next minimum, was identified as the disassembly interval. Several disassembly intervals are indicated in Fig.~2 with thicker red lines.   Between $130$ and $1300$ oscillations were analyzed for every parameter set, depending upon the period and the computational efficiency. 

Statistics of stuttering, polymer number, and length were extracted only from disassembly intervals. Polymers of MinD were counted if they were of length two or more, i.e. monomers were excluded. Stuttering was defined by a transient increase of polar MinD during the disassembly interval, and the duration of the transient increase was the stutter duration.  While excluding short stutters reduced the number of stutters, and degraded their statistics, it did not appear to change their overall functional dependence on various model parameters. Accordingly, we typically counted any stutter that lasted for $1s$ or more --- accessible to experimental timescales and avoiding fluctuations due to individual polymerization events.  The stutter rate was defined to be the average number of stutters observed per disassembly interval.

%%%%%%%%%%%%%%%%%%%%%%%%
%%%%%%%%%%%%%%%%%%%%%%%%%
\section {Results}

We recovered pole-to-pole Min oscillations for a wide range of parameters. The oscillation period was strongly dependent (data not shown) on the ratio of MinD to MinE copy numbers, with a critical minimum ratio required to sustain oscillations  \cite{Raskin1999, Huang2003}.  We found oscillations with both helically pitched (data not shown) and with straight MinD filaments,  with both filament fragmentation and with tip-directed depolymerization, and with a variety of bacterial lengths and widths. In all cases the oscillations were end-to-end, and were observed in both length $L=2 \mu m$ and $L=4 \mu m$ cells. Typically the shorter cells stuttered more. With some parameter sets we observed oscillations only with heterogeneous binding. For simplicity, and in lieu of direct biophysical measurements of most of the interaction parameters, we have restricted ourselves to one core parameter set (see caption of Fig.~2) that exhibits oscillations with both homogeneous and inhomogeneous PL patches, one bacterial size,  and with axially-oriented MinD filaments. 

Fig.~2 shows spatio-temporal Min oscillations in a cell of length 4 micron cell through the process of periodic growth and decay of several polymer filaments at alternate ends of the cell. Filament nucleation was restricted to the hemispherical poles and MinD monomer binding was enhanced there as well-- this is motivated by the inhomogeneous CL distribution seen {\em in vivo}. For the sake of clarity, only  membrane-bound Min molecules are shown in the figure with blue representing MinD-ATP and red representing MinE bound to MinD-ATP. MinD filament scan be observed to start forming in the left end of the cell in Fig.~2(a). As time progresses, the filaments gradually grow longer and become more numerous (panels (b) and (c)) and many of them are also decorated by MinE, though the MinE are still sparsely distributed in panels (b) and (c). The gradual shrinking of the filaments that are predominantly capped by MinE (the E-ring) can also be observed in the opposite (right) end of the cell, in synchrony with the growth of MinD-ATP filaments in the left end of the cell. The growth of the filaments on the left end is eventually halted and shrinking of filaments, primarily by the MinE stimulated hydrolysis of MinD-ATP, is observed in panels (d), (e) and (f).  This E-ring driven shrinking is associated with growth of the MinD filaments at the opposite pole. The period of oscillations was approximately 42 seconds, which is consistent with observations in WT cells at room temperature.  

In Fig.~3 we consider the same set of parameters but with homogeneous nucleation and binding of MinD along the membrane (i.e. $\sigma_{Dcl}=\sigma_D$) --- corresponding to the absence of anionic PL patches at the cell poles. Panels (a-f) show snapshots of the resulting oscillation; the filaments are more uniformly distributed along the length of the cell and are both shorter and more numerous. The oscillation period is approximately double the inhomogeneous case. 

%%%%%%
\subsection{Effects of $P_{protect}$ and $P_{passE}$ on the Min oscillation}
In Fig.~4 we consider the effects of relaxing either $P_{passE}$ or $P_{protect}$ from $P_{passE}=P_{protect}=1$, where we always have processivity of released MinE from filament tips and always protect MinE-bound tips from growth. As either $P_{protect}$ (red squares) or $P_{passE}$ (blue stars) decrease, oscillation periods grow both longer and more variable. The effect is more pronounced as $P_{protect}$ is decreased. In the inset, we see that the amplitude of the oscillation, as measured by twice the standard deviation of the polar MinD content ($2 \sigma_D$), remains appreciable over the parameter ranges shown.  We note that regular oscillations were not observed with no processivity ($P_{passE}=0$), and that (data not shown) shorter $L=2 \mu m$ cells had appreciably longer and more variable periods for $P_{passE}<1$ --- which was not seen with any other parameter variation in this figure or subsequent.

Qualitatively, both $P_{passE}$ and $P_{protect}$ improve the quality of oscillation by strengthening the decoration of filament tips with MinE. With a fixed microscopic stimulated disassembly rate $k_S$, oscillations are slowed with decreased $P_{passE}$ or $P_{protect}$ by interrupting the rapid disassembly of MinD filaments. Interestingly, oscillations are much more sensitive to $P_{protect}$ than to $P_{passE}$ because of the approximately uniform concentration of $MinD-ATP$ monomers ready to assemble unprotected filaments. The interruptions are stochastic, and lead to an increased variability of the cycle-to-cycle duration. We expect that this will also be associated with an increased stutter rate.

In Fig.~5 we consider the stutter rate, as measured by the number of transient reversals of polar disassembly per oscillation, for the same parameter ranges as in Fig.~4.  We show stutters of longer than $1s$, $2s$, and $3s$ with solid, dashed, and dotted lines respectively. We see that the functional form of the different stutter durations are similar, and we subsequently show all stutters of $1s$ or longer duration. We also see that very few stutters are observed until $P_{passE} \lesssim 0.8$, while significant stutters are seen for any $P_{protect}<1$.  We note that no more than about $1000$ oscillation periods were recorded for any parameter set, so stutter rates below $10^{-3}$ were not observable.  In the inset we show the average filament length during the disassembly phase, measured in monomers, at corresponding values of $P_{protect}$ (red squares) or $P_{passE}$ (blue stars). Reduce stuttering corresponds to shorter filaments, on average. 

Tip protection, through $P_{protect}$, and processivity, through $P_{passE}$, both reduce tip-poisoning and hence stuttering.  This echoes what was seen in the previous figure with the degradation of the oscillation period.  Better MinE coverage of filament tips during disassembly leads to quicker disassembly, shorter periods, shorter filaments, less stochastic pausing during disassembly, and less stuttering.  Both $P_{protect}$ and $P_{passE}$ must be close to unity for reliable oscillations with infrequent stuttering.

%%%%%%
\subsection{Effects of filament cutting ($k_f$) on the Min oscillation}
In the previous section we explored models with unbroken MinD filaments by only allowing tip-directed disassembly. In this section we allow MinE-stimulated ATPase activity to break filaments away from from MinD filament tips through the cutting or fragmentation rate $k_f$. As shown in Fig.~6, small levels of filament cutting significantly decreases stuttering even in systems (green circles) with no processivity or tip-protection. As shown in the inset, the corresponding oscillations are regular with large amplitudes. Small levels of cutting also reduces stuttering in conjunction with processivity (purple squares) and with partial protection and processivity (orange triangles).  We see a broad minimum of the stutter rate near $k_f \approx 0.1/sec$, and the subsequent increase of apparent stuttering is associated with (see inset) a significant reduction of the oscillation amplitude. Qualitatively, when $k_f \gtrsim 0.5/sec$ the traces for polar MinD appear (not shown) to be noisy and irregular. Nevertheless, for $k_f \lesssim 0.2/sec$ good oscillations are observed for all of the systems investigated. 

As shown in Fig.~7, increasing $k_f$ leads to shorter MinD filaments (solid lines, same colour and point type as Fig.~6) and more MinD filaments (dashed lines). As shown in the inset, this also leads to shorter oscillation periods --- following from faster disassembly of the shorter filaments. Further increasing $k_f$ continues the trend. For our model parameters, regular oscillations are not observed with  filaments that are shorter than approximately $10$ monomers. 

Qualitatively, excessively high fragmentation ($k_f \gtrsim 0.1/sec$) degrades the oscillation amplitude. Microscopic stochastic effects, such as transient polymerization of individual filaments or nucleation of new filaments, are then more likely to lead to transient increase of the total MinD (i.e. stuttering, as we have measured it). This results in the increase of measured stuttering at larger $k_f \gtrsim 0.1/sec$, and a broad minimum of the stuttering rate. Interestingly, the minimal stuttering rate without any tip-protection or processivity (green circles) is significantly increased if processivity ($P_{passE}=1$) is added without tip-protection (purple squares). How could processivity {\em increase} stuttering?   We speculate that in this case processivity leads to some filaments retaining more bound MinE as they rapidly disassemble --- allowing other filaments to experience increased tip poisoning which then leads to increased collective stuttering.  This increase is avoided when tip-protection is also added (orange triangles). Clearly, there is a rich interplay between multiple filament tips mediated by the association dynamics of MinE and cytoplasmic MinD.

%%%%%%%%%%%%%%%%%%%
%%%%%%%%%%%%%%%%%%%
\section{Discussion and conclusions}

We have demonstrated a polymeric model of Min oscillations with MinD filaments, consistent with structures seen both in vivo \cite{Shih2003} and in vitro \cite{Hu2002, Suefuji2002}. Like the polymeric Min model of Pavin {\em et al} \cite{Pavin2006, Krstic2007}, our model is 3D and fully stochastic, including random motion of individual cytoplasmic proteins within the bacterial volume. However, we also systematically explore stuttering --- the transient reversal of polar MinD disassembly that has been observed in vivo \cite{Shih2002EMBO, Hale2001EMBO, Hsieh2010}.   Stuttering has not been recovered in any non-polymeric model of Min oscillations, and we believe it to be intrinsically a polymeric phenotype \cite{Fange-Elf2006}. 

Frequent poisoning (i.e. binding of MinD to a DE-complex at the tip of a polymer filament) of a rapidly depolymerizing MinD filament by denuding the filament tip of MinE is an issue facing normal Min oscillations in quantitative polymeric models.  The existence within the cell of filament tips that are polymerizing at the same time that others are depolymerizing means that a depolymerizing MinD tip that is poisoned will rapidly switch to polymerization, slowing the oscillation and causing a stutter.

To recover fast regular oscillations, we needed to suppress stuttering. Three mechanisms in our model reduced stuttering. The first was MinE processivity, the immediate rebinding of MinE associated with MinD filament tips upon MinD depolymerization --- through $P_{passE}$. The second was to protect MinD from binding to the tip of an already depolymerizing (MinE decorated) MinD filament --- through $P_{protect}$.  We found that both of these mechanisms together were sufficient to recover fast oscillations with long filaments, as illustrated in Fig.~2 with $P_{passE}=0.9$ and $P_{protect}=1$. However, oscillations were more sensitive to changes in $P_{protect}$ than $P_{passE}$ with oscillations degenerating even for a value of $P_{protect}$ as large as $\approx 0.80$. Independently of processivity and tip-protection, we were also able to suppress stuttering by allowing MinE bound away from filament tips to cut MinD filaments. This mechanism has been included in previous polymeric Min models \cite{Pavin2006, Krstic2007, Tostevin2006}, but was not systematically explored.  

In general we found that more stuttering is associated with longer oscillation periods and longer filaments. For parameter values that recovered typically observed oscillation periods in the range of 10-100s a stutter rate of approximately one stutter in every 100 oscillations was observed.  For some parameter values no stutters were observed in hundreds of oscillations.   Combining several stutter suppression mechanisms, such as filament cutting and protecting filament tips from poisoning, led to lower stuttering rates.  

How does this compare with stuttering observed {\em in vivo}? WT oscillations  ``occasionally'' stutter \cite{Hale2001EMBO}, which we take as no more than 1\% stutter per cycle. The ``C1'' MinE mutant (R10G/K11E/K12E) \cite{Hsieh2010} exhibited frequent stuttering,  extended MinD polar zones,  slower oscillations, and weaker E-rings. The D45A/V49A MinE mutant was qualitatively similar \cite{Shih2002EMBO, Hsieh2010}. The quantification is crude, but oscillation periods of 2-3x slower and stutter rates of approx 50\% of cells over 1 hour (corresponding to approximately 10\% per cycle) are consistent with {\em all} of our mechanisms.  However, the observation of longer MinD polar zones in the stutter mutants \cite{Shih2002EMBO,Hsieh2010} together with the infrequent reports of long filaments in WT Min oscillations (only \cite{Shih2003,Szeto2005}), could both be explained by significant filament fragmentation operating in WT oscillations, and reduced fragmentation and longer MinD filaments in the stutter mutants.  

While more structural evidence is accumulating on how MinE binds to MinD \cite{Park2011, Wu2011}, there are essentially no measurements of biochemical rates or of MinD polymer lengths that might distinguish our mechanisms. We note that our processivity mechanism is qualitatively similar to the ``Tarzan of the jungle'' model of Park et al \cite{Park2011}, in which active MinE is passed from one membrane associated MinD to another. We do note that such processivity is likely to be local, and so may be limited to polymeric models: even very slow protein conformational timescales of $1ms$ would only allow several $nm$ of diffusion --- much less than the expected separation of isolated MinD on the membrane but not so different from monomer spacing within MinD polymers. Hence, we believe processivity is inherently polymeric.

The observed minimal stutter rate of less than 1\% per cycle at intermediate filament cutting rates ($k_f \in [0.01, 0.1]/sec$,  see Fig.~\ref{FIGstutterfrag}), corresponding to short MinD polymers with length between 10 and 20 monomers (see Fig.~\ref{FIGnumpolyfrag}) leads us to believe that a polymeric model with filament cutting is appropriate for describing WT Min oscillations {\em in vivo} \cite{ Pavin2006, Krstic2007, Tostevin2006}. The expected filament cutting rate, $k_f$, depends on other mechanisms that can suppress stuttering, such as processivity $P_{passE}$ \cite{Park2011} and tip-protection $P_{protect}$.

It would be interesting to compare our results with stutter rates in other stochastic polymeric models of Min oscillations \cite{ Pavin2006, Krstic2007, Tostevin2006} as fragmentation rates are varied.  We anticipate that 1D models \cite{Tostevin2006} may effectively enhance the true processivity $P_{passE}$ due to recurrence of random walks in one-dimension --- so studies in 3D are called for. Unfortunately 3D molecular-dynamics simulations such as our study, and those of Pavin {\em et al} \cite{Pavin2006, Krstic2007} are slow --- and extensive parameter searches to achieve quantitative agreement with experimental phenomenology are not practical. Given the variety of mechanisms that are needed to restrict stuttering, and the number of other parameters in Min oscillation models, what is needed is careful experimental characterization of microscopic rates and structures akin to what is available for actin polymerization \cite{Pollard1986}. We expect that this is possible in {\em in vitro} systems \cite{Loose2008}.  

\ack
This work was supported financially by the Natural Sciences and Engineering Research Council (NSERC), and with fellowship support for JD from the Atlantic Computational Excellence Network (ACEnet). Computational resources were provided by ACEnet. SS would also like to acknowledge use of the high performance computing facility, at Jawaharlal Nehru University, that was funded by the University Grants Commission (UGC), India through an UPOE grant.  
\newpage

%%%%%%%%%%%%%%%%%%%%%%%
%%%%%%%%%%%%%%%%%%%%%%%

%%%%%%%%%%%%%%%%%%%%%%%%%%%%%%%%%%%%%%%%%
%%%%%%%  FIGURES %%%%%%%%%%%%%%%%%%%%%%%%%%%%

%%%%%%%%%%%%%%%FIG 1 %%%%%%%%%%%%%%%%%%%%%%%
\newpage
\begin{figure}
\centering{
\includegraphics[width=0.7\textwidth,clip] {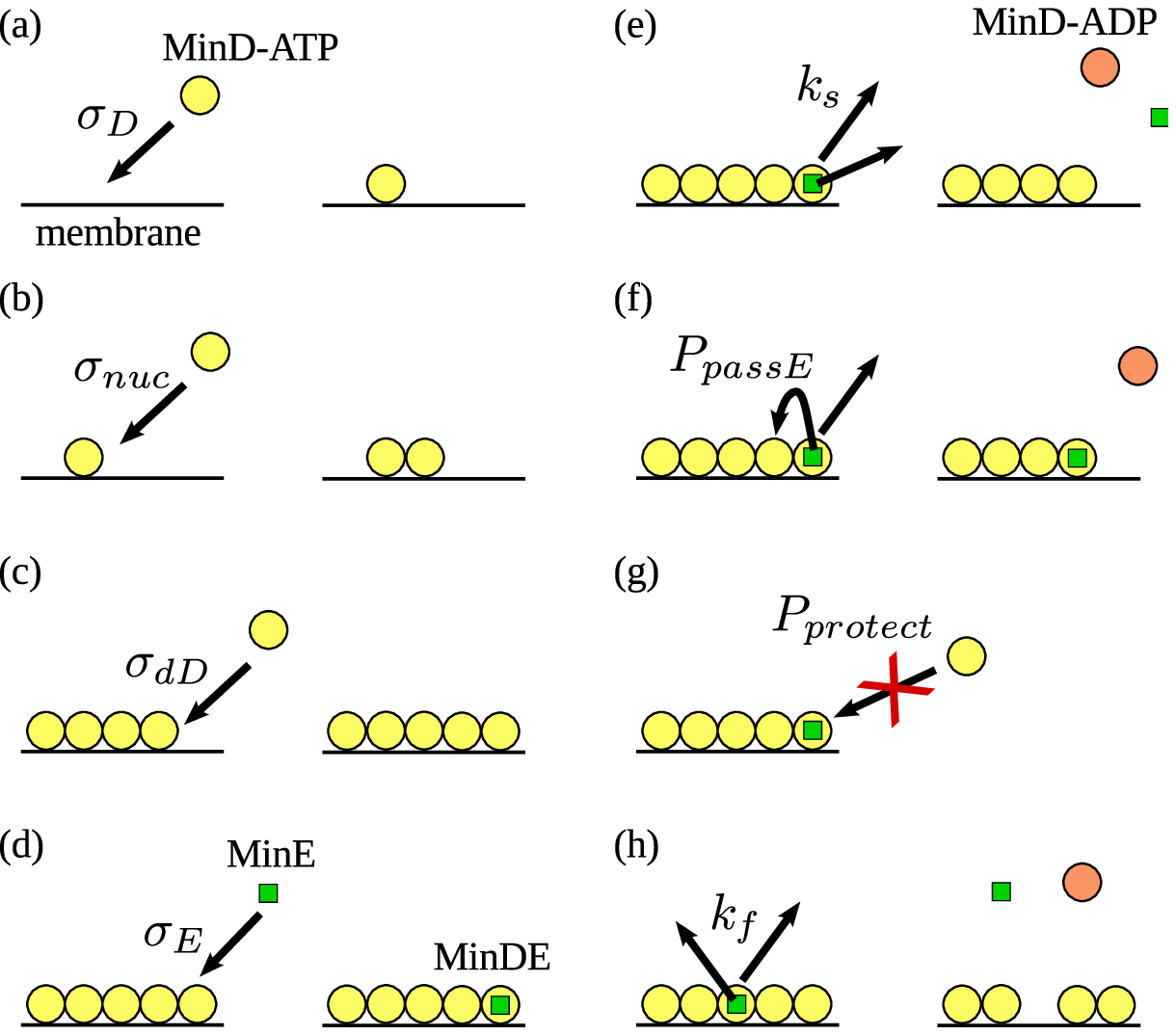}
} % centering
\caption{{\bf Schematic of kinetic parameters in our Min model} 
(a) Cytoplasmic MinD-ATP (yellow circles) binds to the membrane with rate proportional to $\sigma_D$; (b) Filaments nucleate by binding of second MinD-ATP to a membrane-associated MinD-ATP, with rate proportional to $\sigma_{nuc}$;  (c) Filaments elongate by tip-binding of MinD-ATP, with rate proportional to $\sigma_{dD}$; (d) MinE (green square) binds to {\em any} of the filamentous MinD (binding to the tip is illustrated) with rate proportional to $\sigma_E$ , leading to MinDE);  (e) MinE stimulated MinD-ATPase activity of MinDE leads to disassociation of MinD-ADP (darker, rose coloured, circle) and MinE from the filament tip, with rate $k_S$; (f) With probability $P_{passE}$ the tip-released MinE from (e) will processively associate with an adjacent MinD-ATP rather than being released into the cytoplasm; (g) With probability $P_{protect}$, MinD-ATP binding will be blocked from filament tips that have MinDE; (h) With rate $k_f$, MinDE away from filament tips will disassociate and fragment the filament into two. We find that the last three processes (f-h) control stochastic stuttering of Min oscillations.
}
\label{fig1}
\end{figure}

%%%%%%%%%%%%%%%FIG 2%%%%%%%%%%%%%%%%%%%%%%%
\newpage
\begin{figure}\begin{minipage}{\textwidth}
\centering{
\mbox{
\subfigure[]{\includegraphics[width=.3\textwidth]{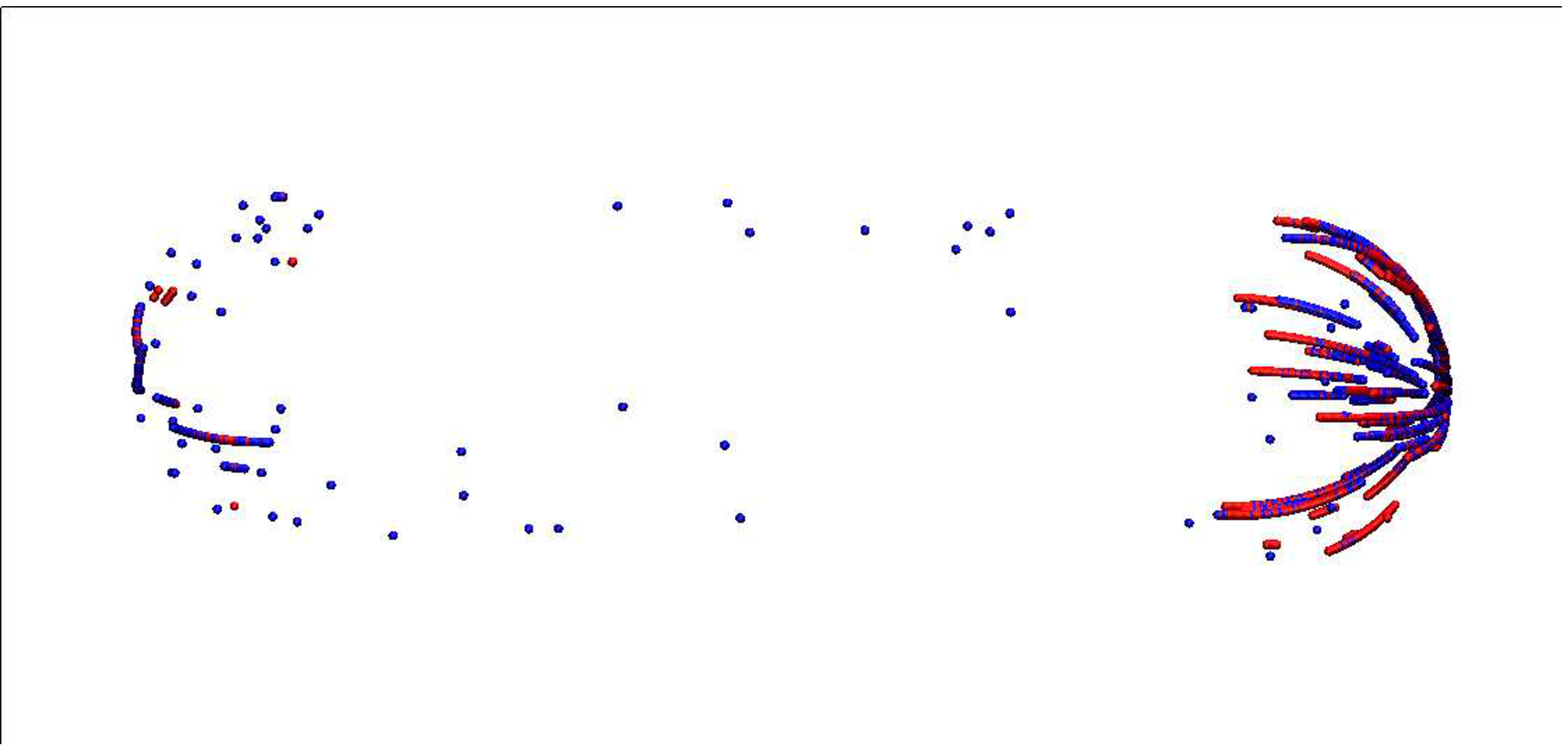}\label{(2a)}}
\subfigure[]{\includegraphics[width=.3\textwidth]{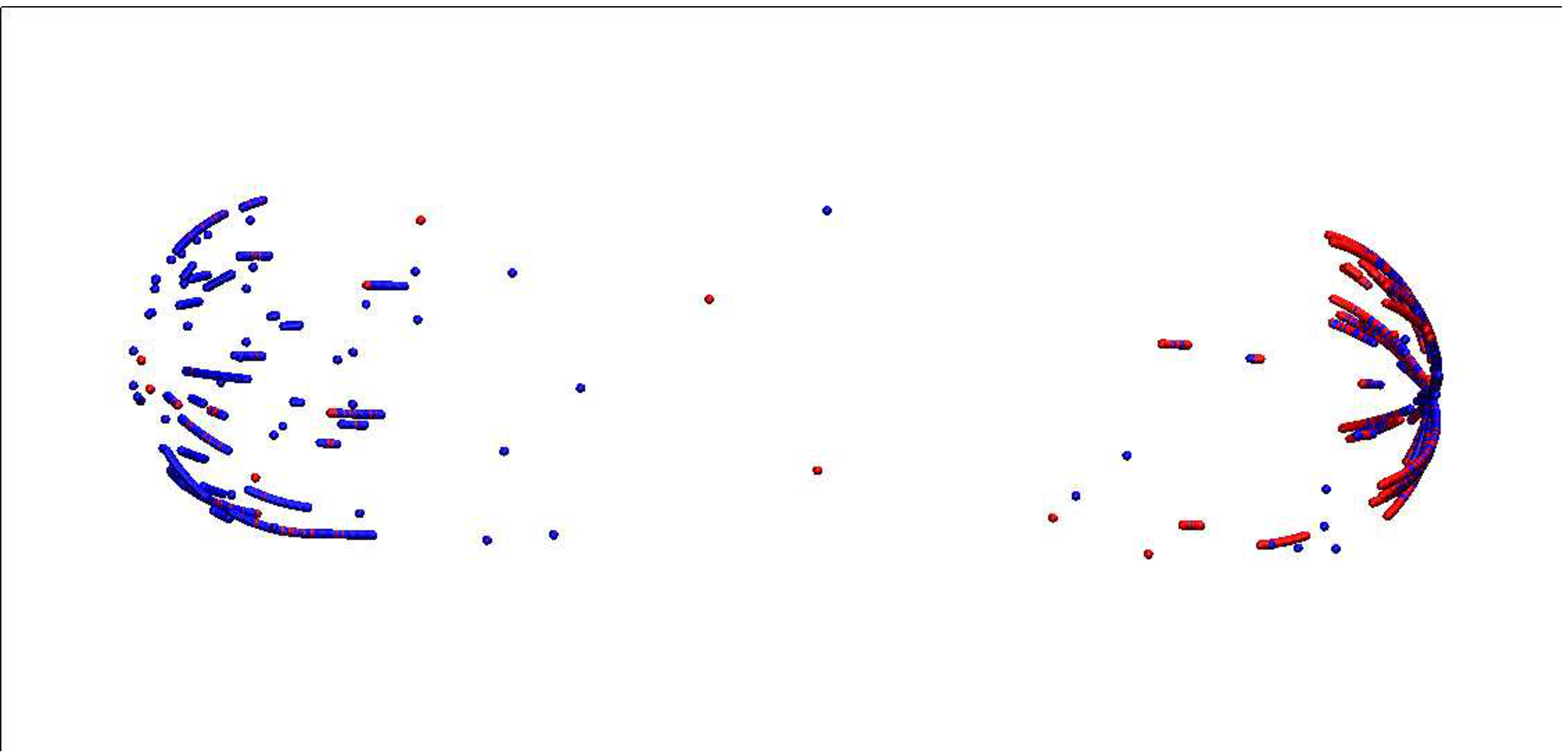}\label{(2b)}}
\subfigure[]{\includegraphics[width=.3\textwidth]{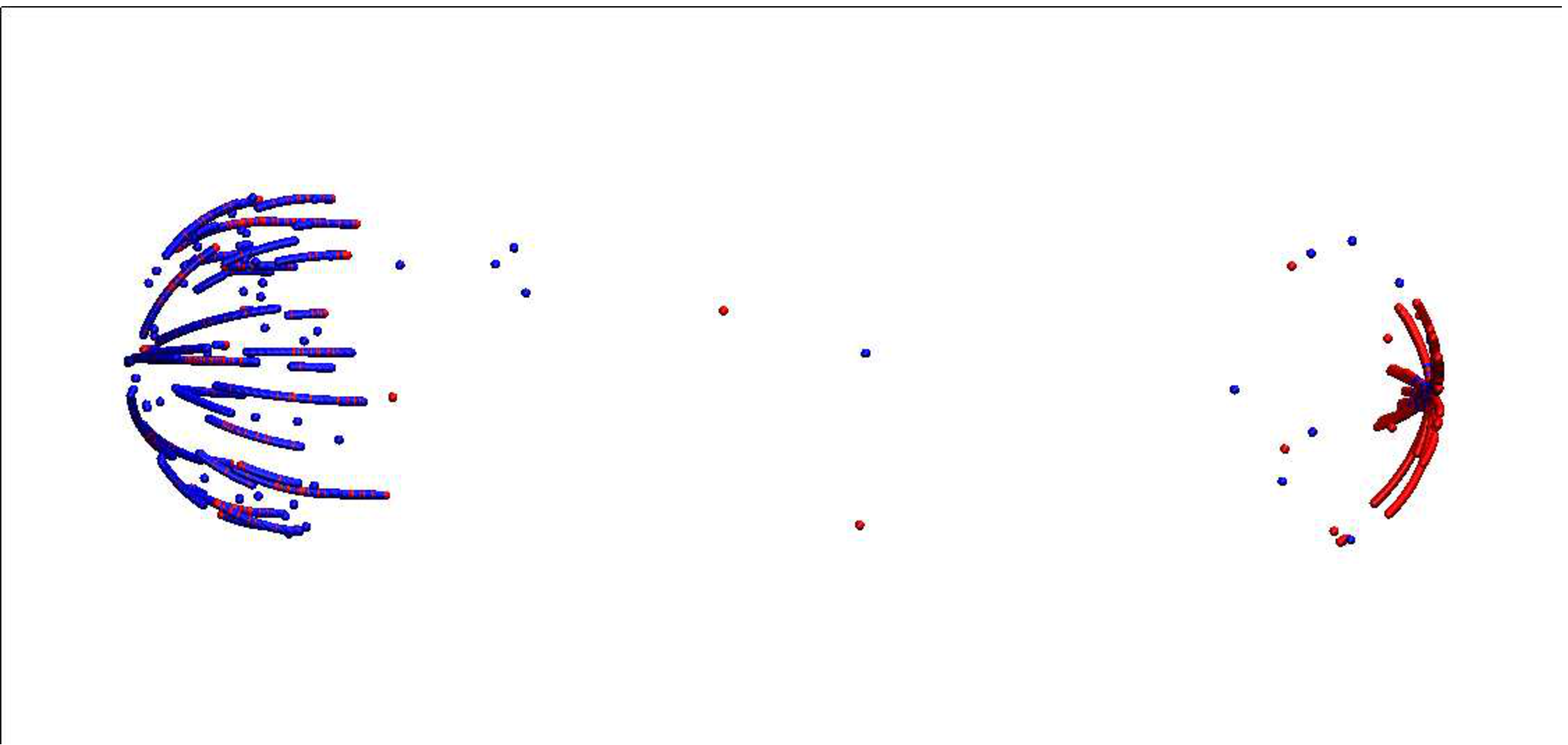}\label{(2c)}
}\quad}
\mbox{
\subfigure[]{\includegraphics[width=.3\textwidth]{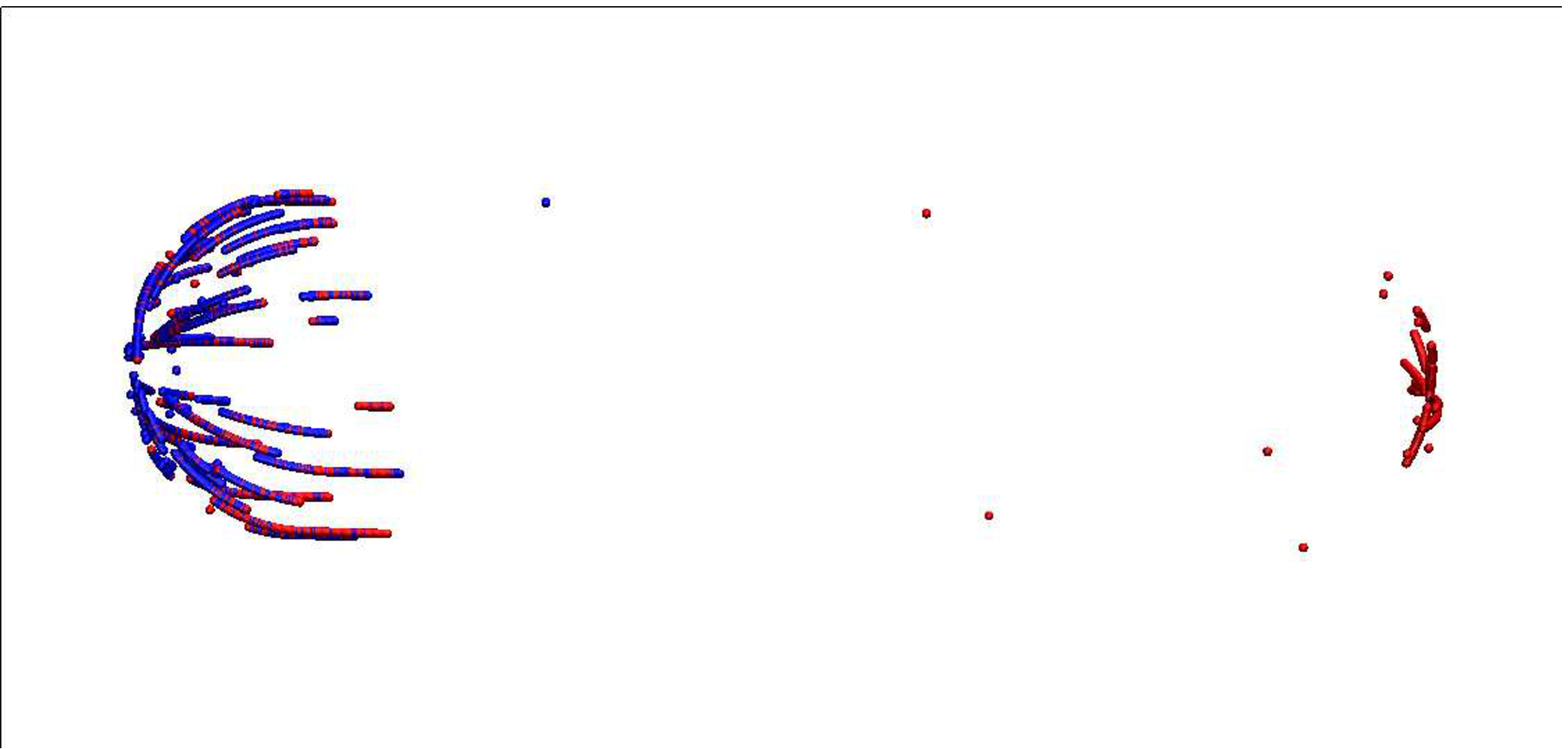}\label{(2d)}}
\subfigure[]{\includegraphics[width=.3\textwidth]{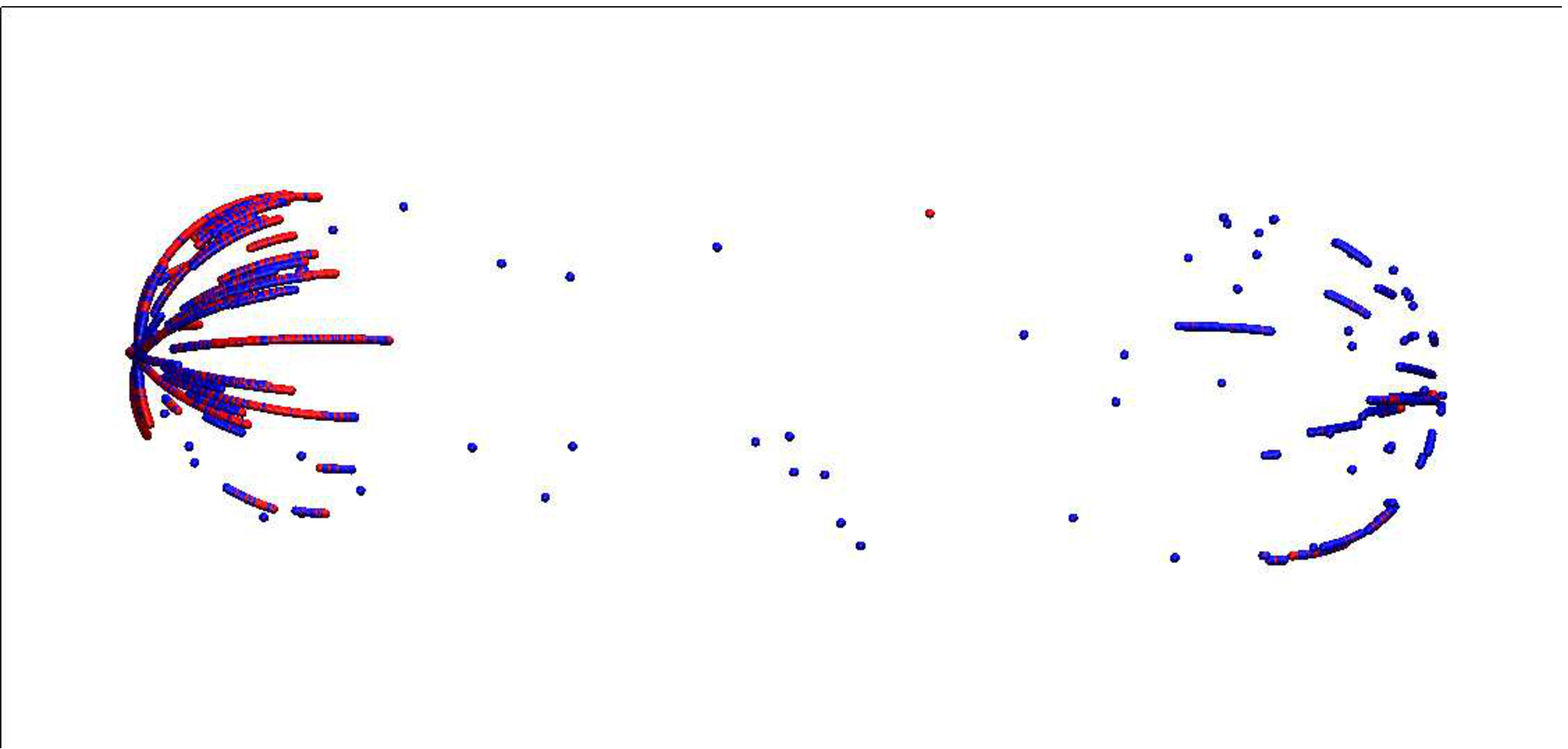}\label{(2e)}}
\subfigure[]{\includegraphics[width=.3\textwidth]{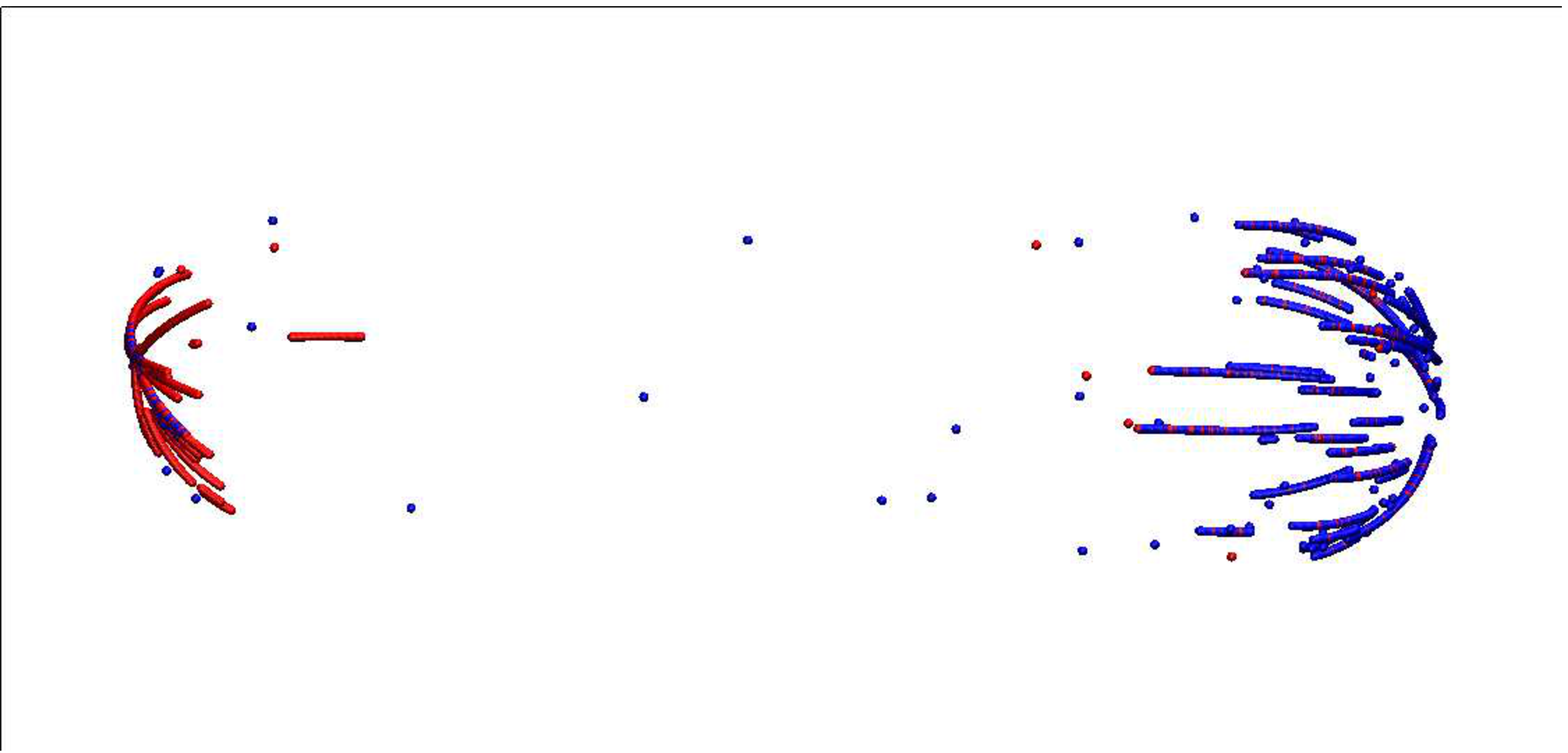}\label{(2f)}}\quad}
\mbox{\subfigure[]{\includegraphics[angle=270,width=.7\textwidth]{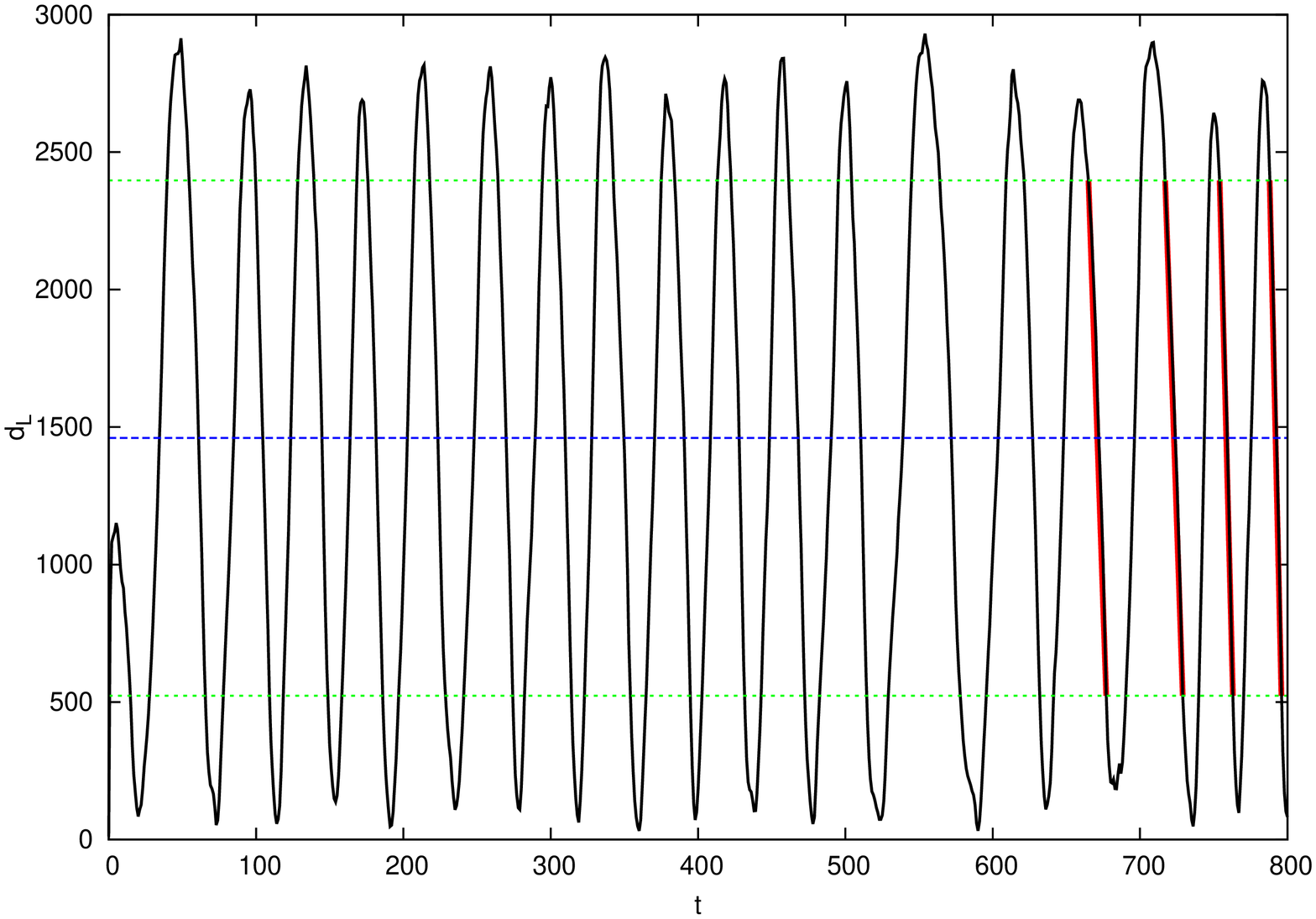}\label{(2g)}}}
}
\caption{{\bf Min oscillations - inhomogeneous nucleation.} Consecutive snapshots (a-f) from our simulation video shows periodic polymerization and depolymerization of MinD filaments at the two ends of a rod-shaped cell.  The cylindrical portion of the cell has length $L=4 \mu m$ with radius $0.5\mu$m; the hemispherical end-caps have the same radius.  Only membrane-associated molecules are shown: MinD-ATP in blue and MinD-MinE in red.   We apply inhomogeneous nucleation where MinD filaments can only nucleate at the hemispherical poles, and $\sigma_{Dcl}=10\, \sigma_{D}$. The oscillation of  the number of membrane-bound MinD in the left third of the cell is shown in (g). The oscillation period is approximately $43$ seconds, after a short initial transient from inhomogeneous initial conditions. Also shown are the mean $\langle n_D \rangle$ with a dashed blue line, and $\langle n_D \rangle \pm \sigma_D$ with the two dotted green lines. Four disassembly regions, which go between the dotted green lines, are illustrated with thicker red lines. The parameters are: $\rho_D=1200 \mu m^{-3}$, $\rho_E=400 \mu m^{-3}$, $D_D=16 \mu m^2/sec$, $D_E=10 \mu m^2/sec$, $\sigma_D=100 \mu m/sec$, $\sigma_{dD}=5.5\times10^8 \mu m^3/sec$, $\sigma_E= 8.0\times10^8 \mu m^3/sec$, $\sigma_{nuc}= 1.6\times10^7\mu m^{3}/sec$, $k_I=0.1/sec$, $k_{SM}=10/sec$, $k_S=6/sec$, $k_E=0.01/sec$, $\tau_c =1.0 sec$, $r_D = r_E= 25 nm$, $r_{nuc} = 5 nm$, $\delta t=0.01\, sec$, $P_{passE}=0.9$, $P_{protect}=1.0$.} 
\label{fig2}
\end{minipage}
\end{figure}

%%%%%%%%%%%%%%%Fig 3%%%%%%%%%%%%%%%%%%%%%%%%%%%%%%
\newpage
\begin{figure}\vspace{2pc}
\begin{minipage}{\textwidth}
\centering{\mbox{
\subfigure[]{\includegraphics[width=.3\textwidth]{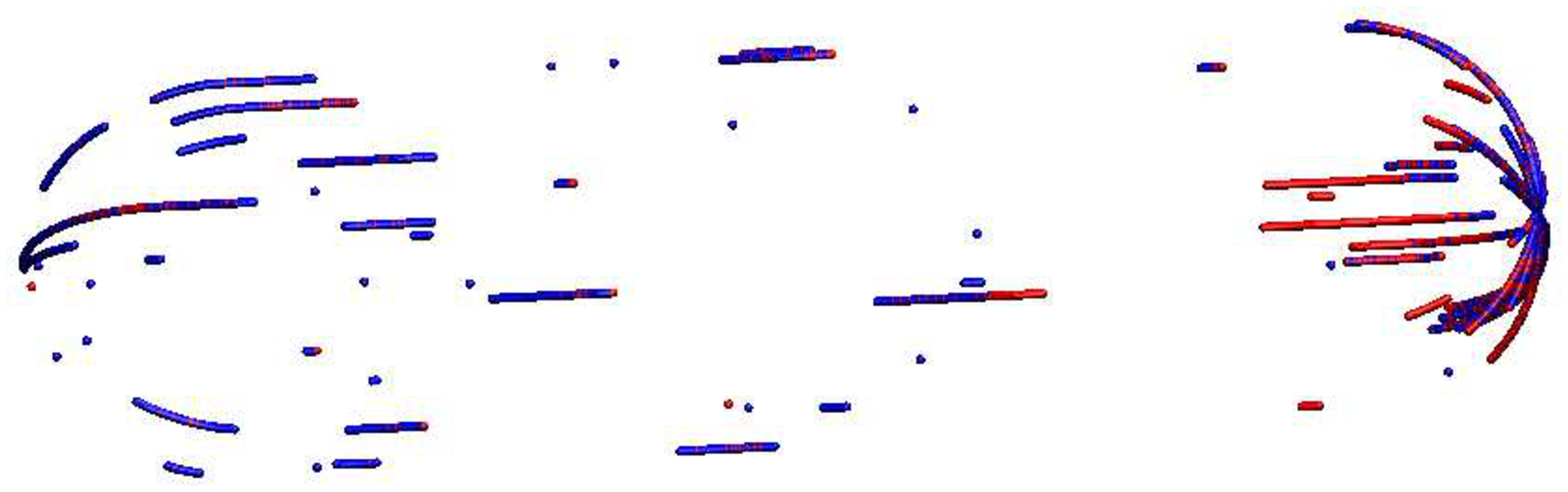}\label{(3a)}}
\subfigure[]{\includegraphics[width=.3\textwidth]{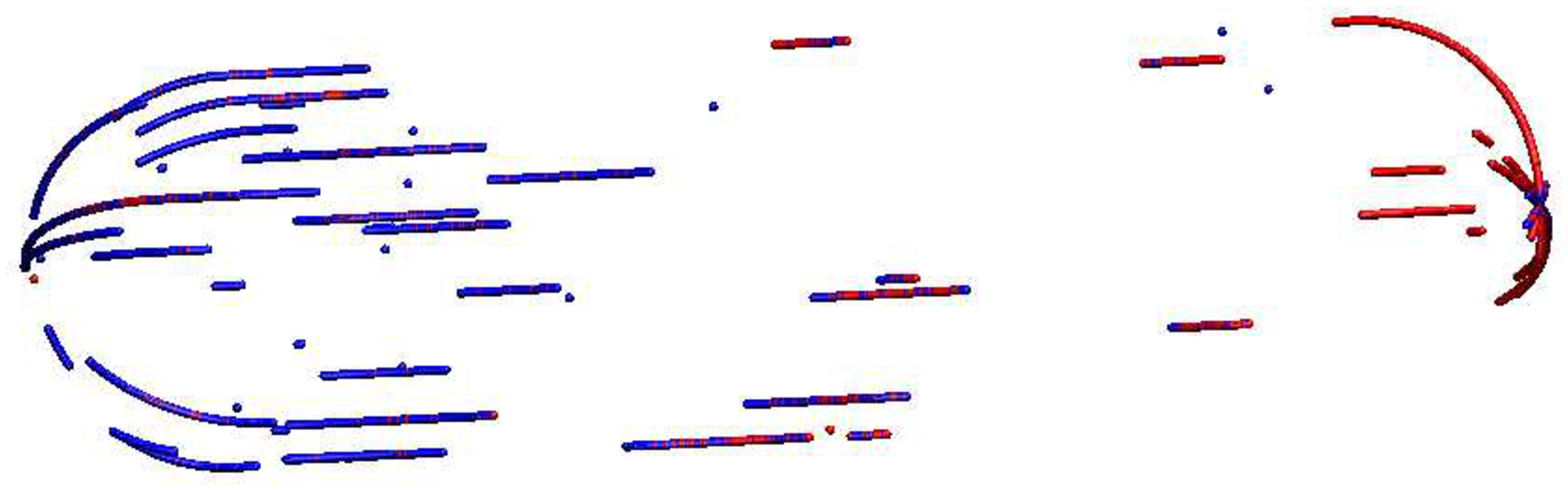}\label{(3b)}}
\subfigure[]{\includegraphics[width=.3\textwidth]{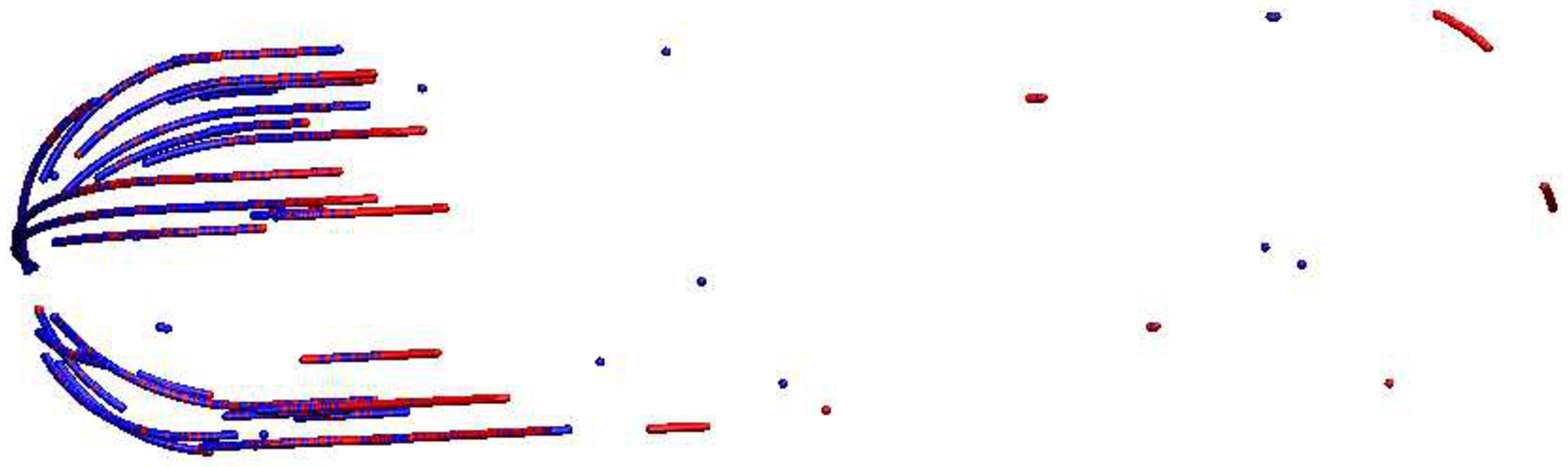}\label{(3c)}
}\quad}
\mbox{
\subfigure[]{\includegraphics[width=.3\textwidth]{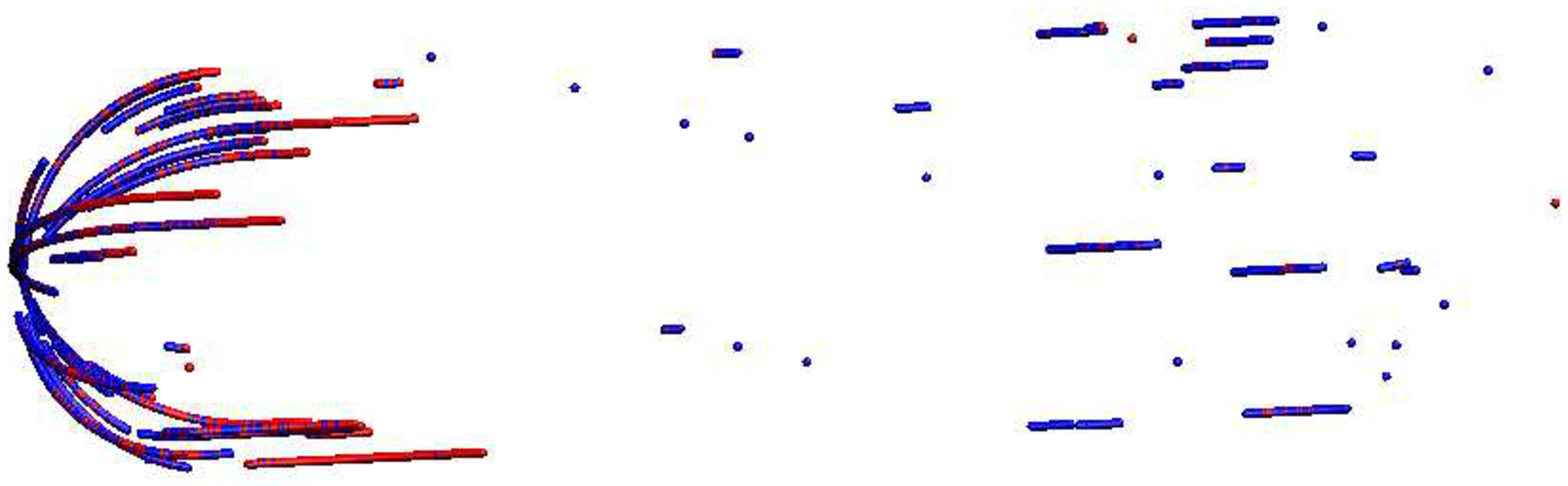}\label{(3d)}}
\subfigure[]{\includegraphics[width=.3\textwidth]{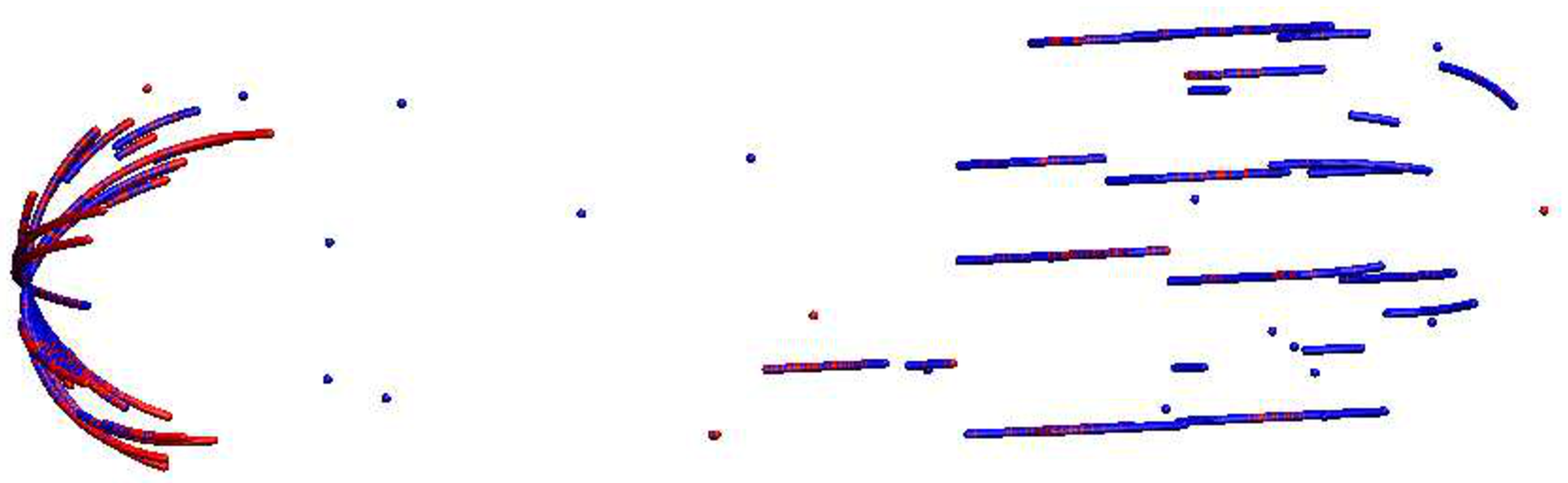}\label{(3e)}}
\subfigure[]{\includegraphics[width=.3\textwidth]{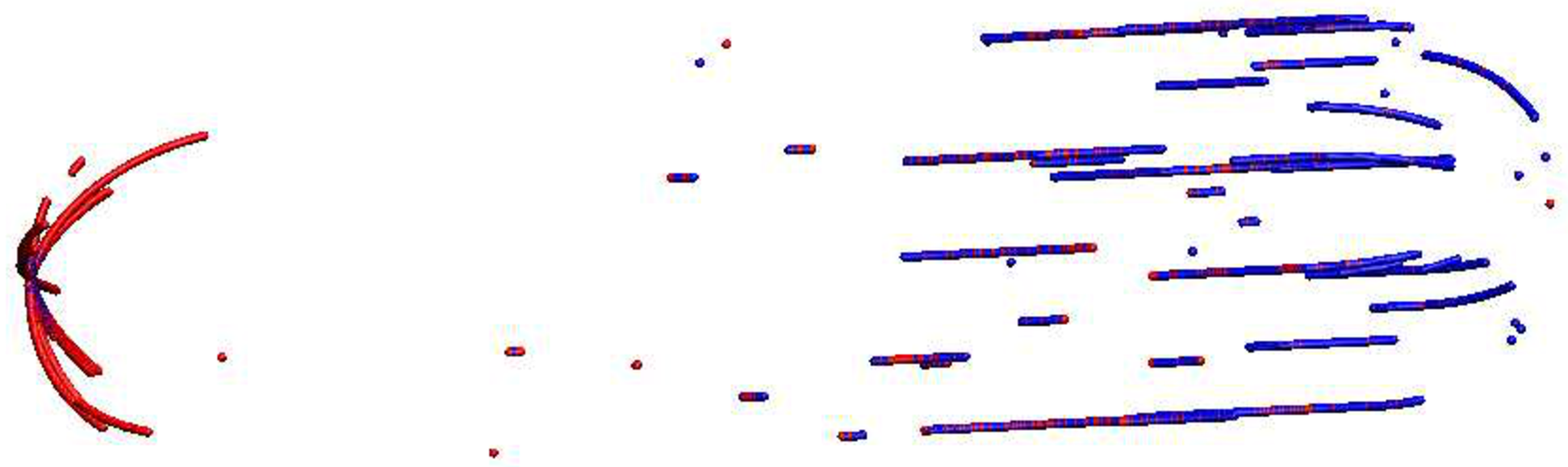}\label{(3f)}
}\quad}
\mbox{
\subfigure[]{\includegraphics[angle=270,width=.7\textwidth]{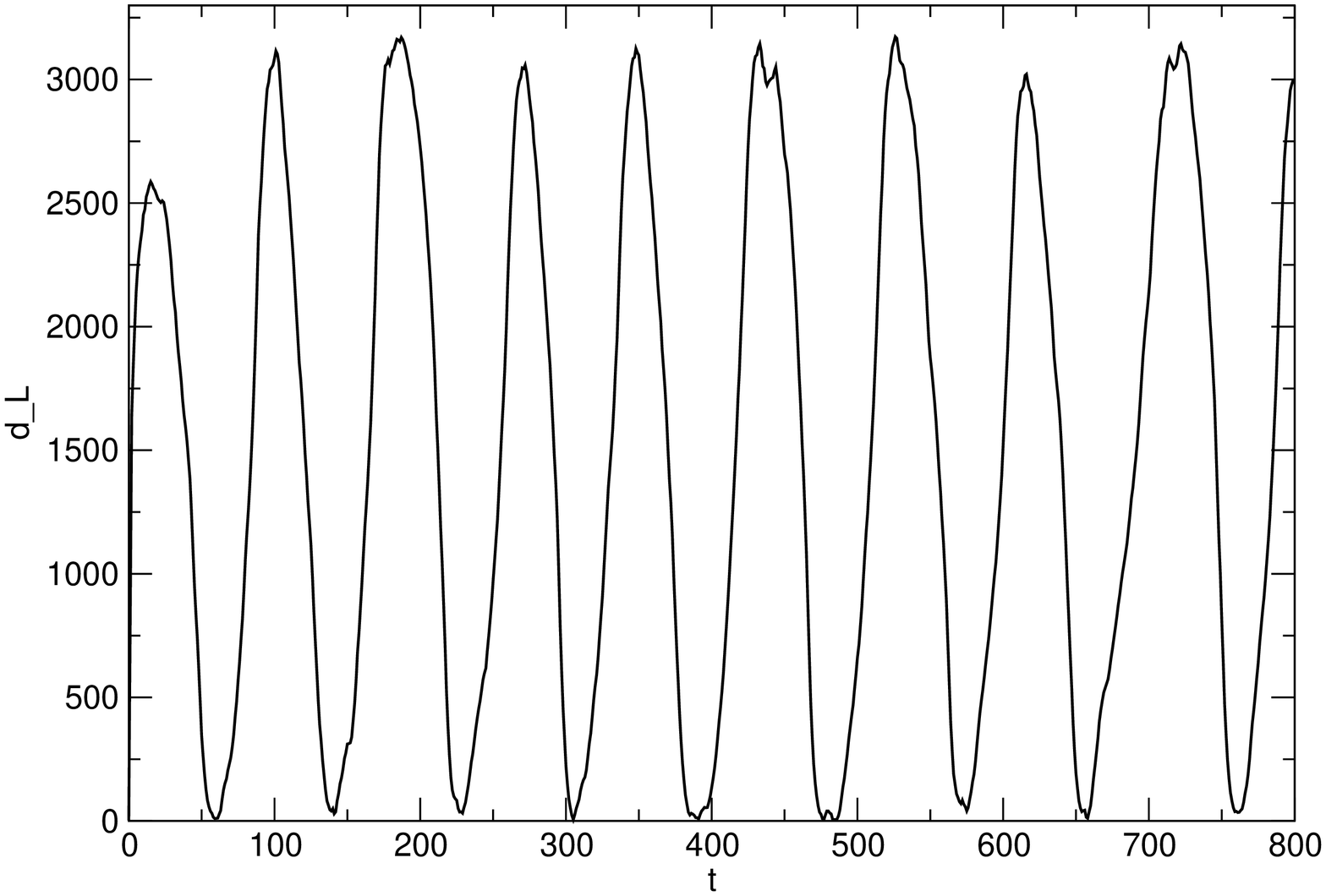}\label{(3g)}}}}
\caption{{\bf Min oscillations - homogeneous nucleation.}
Consecutive snapshots (a-f) from our simulation video shows periodic polymerization and depolymerization of MinD filaments at the two ends of a rod-shaped cell.  Only membrane-associated molecules are shown: MinD-ATP in blue and MinD-MinE in red.   We apply homogeneous nucleation where MinD filaments can nucleate everywhere and MinD monomers bind homogeneously $\sigma_{Dcl}= \sigma_{D}$. Other parameters are the same as in Fig.~2. The oscillation of  the number of membrane-bound MinD in the left third of the cell is shown in (g). The oscillation period is approximately $85$ seconds, and is reached after a short initial transient from inhomogeneous initial conditions. } 
\label{fig3}
\end{minipage}
\end{figure}

%%%%%%%%%%%%%%%Fig  4  %%%%%%%%%%%%%%%%%%%%%
\newpage
\begin{figure}
\centering{ 
\includegraphics[width=0.7\textwidth,clip,angle=-90] {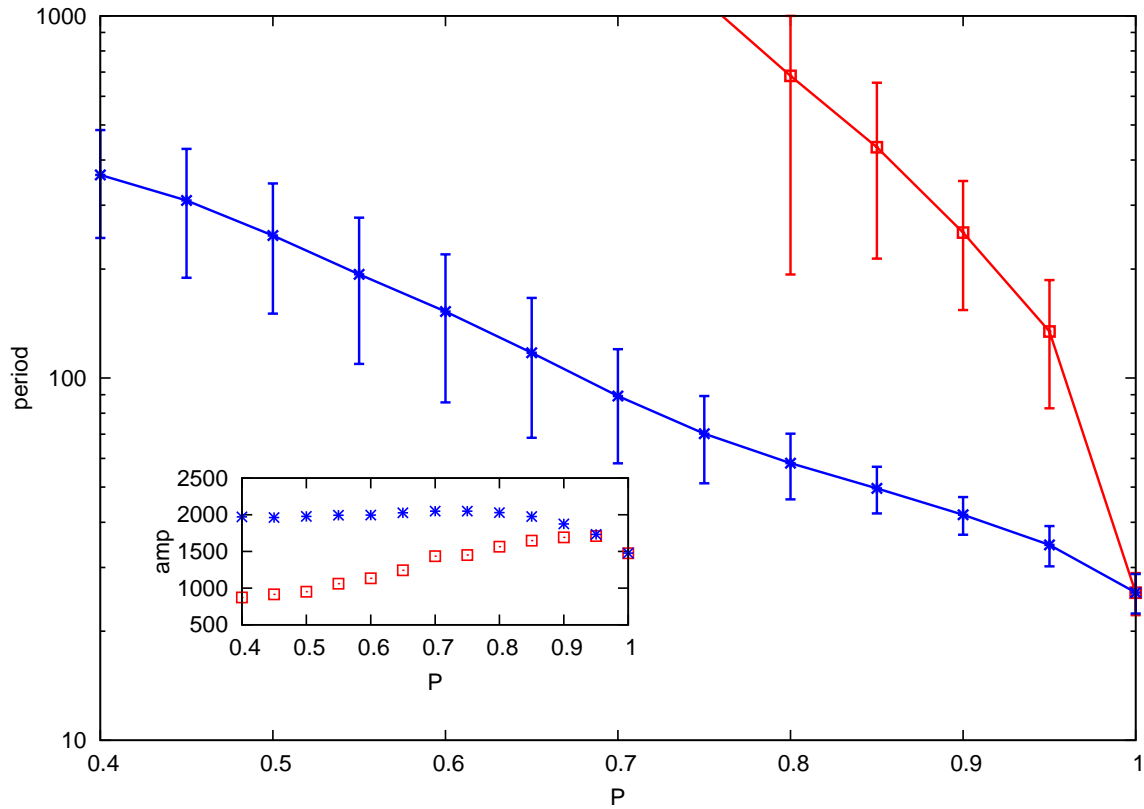}}
\centering{ 
\caption{{\bf Effect of $P_{protect}$  and $P_{passE}$ on the oscillation period.}  We show the mean oscillation period as either tip protection or processivity, $P_{protect}$ (red squares) or $P_{passE}$ (blue stars) respectively, is varied from $P_{protect}=P_{passE}=1$.  The vertical bars indicate the standard deviation of the period, $\sigma_T$ --- statistical errors are much smaller. We see that with maximal processivity and tip protection, oscillations are fast and precise, but that decreasing either $P_{protect}$ or $P_{passE}$ both slows and degrades the oscillation --- this is particularly pronounced with smaller $P_{protect}$. In the inset, we show the corresponding oscillation amplitude ($2 \sigma_D$) vs. $P$. While the oscillation amplitude does degrade somewhat as $P_{protect}$ decreases, the amplitudes remain large for periods less than $1000s$. Parameters for this and subsequent figures are as in Fig.~2, unless otherwise mentioned.}
\label{FIGperiod}}
\end{figure}

%%%%%%%%%%%%%%Fig 5 %%%%%%%%%%%%%%%%%%%%%
\newpage
\begin{figure}
\centering{
\includegraphics[width=0.7\textwidth,clip,angle=-90] {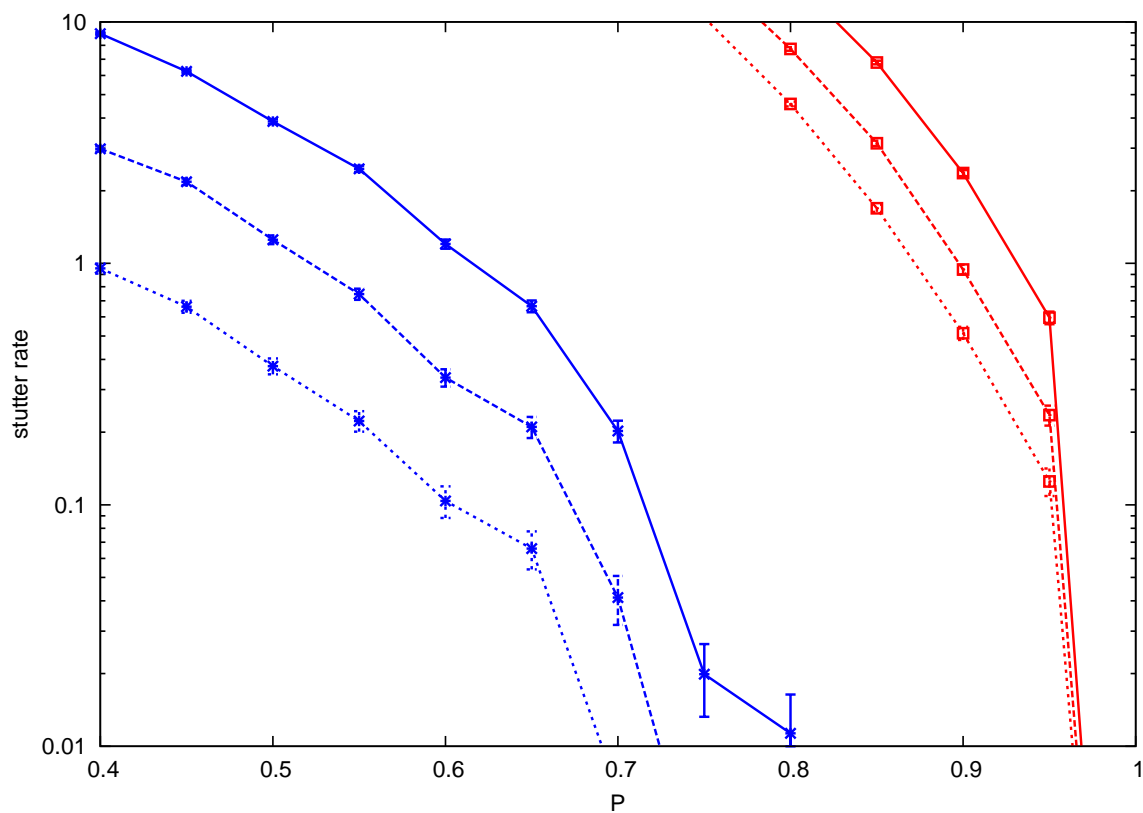}
\caption{{\bf Effect of $P_{protect}$ and $P_{passE}$ on the stutter rate.} The points show the stutter rate per oscillation period as either tip protection or processivity, $P_{protect}$ (red squares) or $P_{passE}$ (blue stars) respectively, is varied from $P_{protect}=P_{passE}=1$.  Statistical error bars are shown. The stutter rate is the average number of transient reversals of the indicated duration ($1s$, $2s$, or $3s$ indicated by solid, dashed, and dotted lines, respectively) or longer during polar disassembly per oscillation.  We see that moving away from full protection or processivity leads to significant rates of stuttering. We also see that stutters of different duration are similarly distributed.  The inset shows the corresponding average filament length, measured in number of monomers.
}
\label{FIGstutterpoisn}}
\end{figure}

%%%%%%%%%%%%%%Fig 6 %%%%%%%%%%%%%%%%%%%%%
\newpage
\begin{figure}
\centering{
\includegraphics[width=0.7\textwidth,clip,angle=-90] {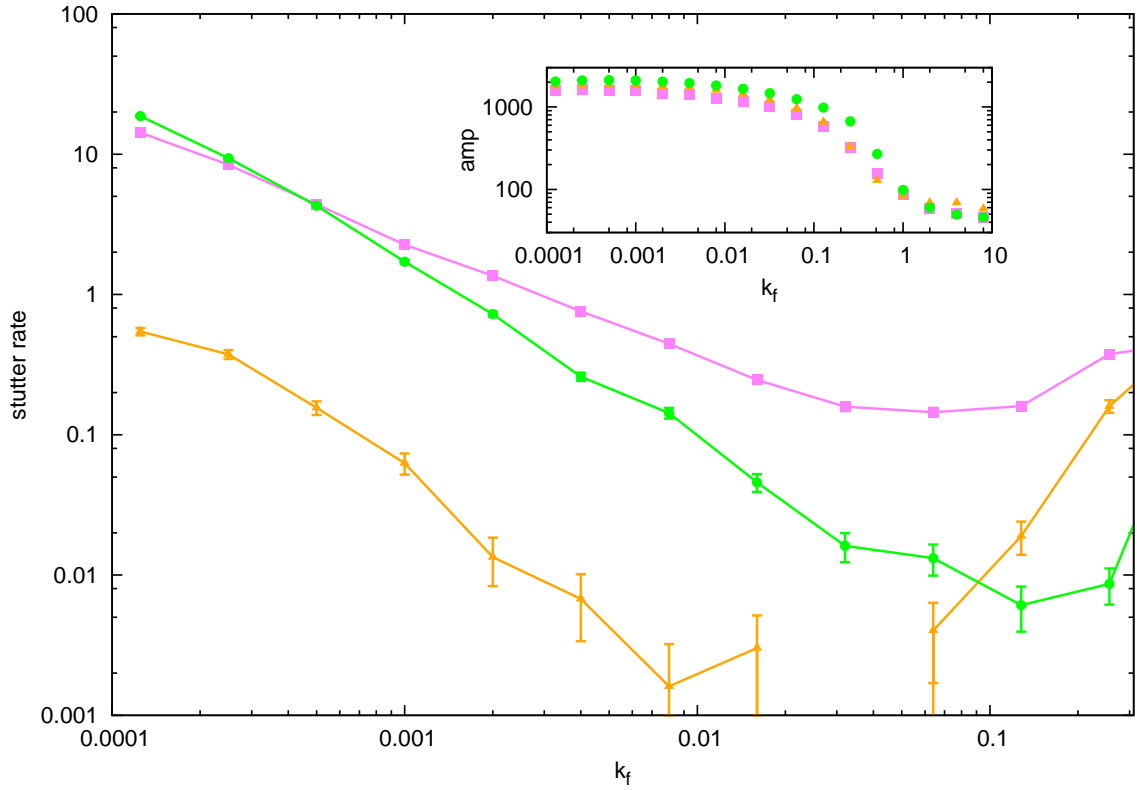}
\caption{{\bf Stutter rate vs fragmentation rate.} Shown is the number of $1s$ or longer stutters per oscillation (the stutter rate) vs. the filament fragmentation rate $k_f$. Statistical errors are as indicated. The various curves are for no protection but full processivity (purple squares), partial protection with $P_{protect}=0.9$ and full processivity (orange triangles), and neither protection nor processivity (green circles). For all three curves, stuttering decreases  as $k_f$ increases from small values, reaches a broad minimum, then increases for further increases of $k_f$. The inset shows the oscillation amplitude ($2 \sigma_D$), which sharply decreases at $k_f \gtrsim 0.1/sec$ --- corresponding to when the stutter rate begins to increase again. Interestingly, processivity significantly increases the minimal stuttering rate (purple vs green curves) unless tip-protection is also present (orange curves). 
}
\label{FIGstutterfrag}}
\end{figure}

%%%%%%%%%%%%%%Fig 7 %%%%%%%%%%%%%%%%%%%%%
\newpage
\begin{figure}
\centering{
\includegraphics[width=0.7\textwidth,clip,angle=-90] {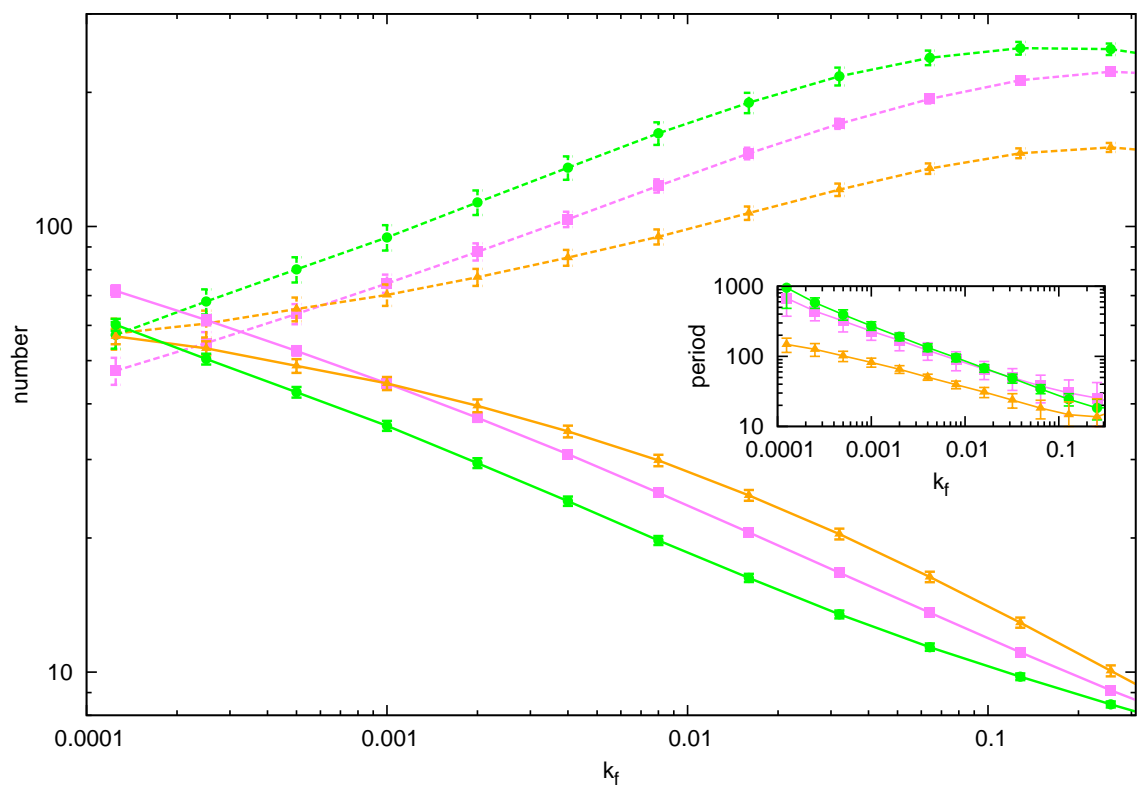}
\caption{{\bf Number of polymers and polymer length (in numbers of monomers) vs fragmentation $k_f$.} The solid lines show the average length of MinD filaments during the disassembly phase of oscillations at either pole, while the dashed lines show the corresponding average number of filaments. The points and colours are the same as the previous figure. The inset shows the corresponding oscillation periods.   We see that increased fragmentation leads to shorter filaments, more filaments, and shorter oscillation periods. 
}
\label{FIGnumpolyfrag}}
\end{figure}

\end{document}